\documentclass[conference]{ieeeconf}
\IEEEoverridecommandlockouts
\usepackage{cite}
\usepackage{amsmath,amssymb,amsfonts}
\usepackage{algorithmic}
\usepackage{graphicx}
\usepackage{textcomp}

\usepackage{comment}

\usepackage{booktabs} 
\usepackage{colortbl} 
\usepackage{xfrac}
\def\Tra{^{\top}}
\usepackage{hyperref}

\begin{document}


\title{Constrained Trajectory Optimization for \linebreak Hybrid Dynamical Systems}

\author{%
Pietro Noah Crestaz$^{1,2}$, Gokhan Alcan$^{3}$, and Ville Kyrki$^{4}$%
\thanks{$^{1}$Industrial Engineering Department, University of Trento, Trento, Italy. \texttt{pietronoah.crestaz@unitn.it}}%
\thanks{$^{2}$LAAS-CNRS, Université de Toulouse, CNRS, Toulouse, France. \texttt{pietro-noah.crestaz@laas.fr}}%
\thanks{$^{3}$Faculty of Engineering and Natural Sciences, Tampere University, Tampere, Finland. \texttt{gokhan.alcan@tuni.fi}}%
\thanks{$^{4}$Department of Electrical Engineering and Automation, Aalto University, Espoo, Finland. \texttt{ville.kyrki@aalto.fi}}%
}


\maketitle

\begin{abstract}
    Hybrid dynamical systems pose significant challenges for effective planning and control, especially when additional constraints such as obstacle avoidance, state boundaries, and actuation limits are present. In this letter, we extend the recently proposed Hybrid iLQR method \cite{kong2021ilqr} to handle state and input constraints within an indirect optimization framework, aiming to preserve computational efficiency and ensure dynamic feasibility. Specifically, we incorporate two constraint handling mechanisms into the Hybrid iLQR: Discrete Barrier State and Augmented Lagrangian methods. Comprehensive simulations across various operational situations are conducted to evaluate and compare the performance of these extended methods in terms of convergence and their ability to handle infeasible starting trajectories. Results indicate that while the Discrete Barrier State approach is more computationally efficient, the Augmented Lagrangian method outperforms it in complex and real-world scenarios with infeasible initial trajectories.
\end{abstract}



\section{Introduction}

In many real-world scenarios, robots such as quadrupeds \cite{Kong2023base}, humanoids \cite{Wensing2023}, and aerial manipulators \cite{ruggiero2018aerial} are tasked with navigating and interacting within complex and dynamic environments. These robotic platforms generally exhibit highly nonlinear dynamics, and their interactions with the environment pose significant challenges to effective planning and control. Often modeled as hybrid dynamical systems, these systems have a discrete component of the state known as the \textit{hybrid mode} across which the continuous dynamics can vary. These modes are linked by a reset function that applies a discrete change to the state.

In addition to their inherent dynamical complexity, these systems often need to meet additional constraints such as obstacle avoidance, state boundaries, and actuation limits. Many control strategies have been specialized for hybrid dynamical systems, as traditional control algorithms for smooth dynamical systems are generally not applicable to hybrid dynamics \cite{kong2023saltation}. Direct optimization methods are the most common ones, which transform the problem into a nonlinear program optimizing both the states and inputs of the system simultaneously. These solutions typically rely on a predefined mode sequence, leading to a parallel optimization of each individual smooth segment connected by boundary conditions \cite{posa2016optimization,schultz2009modeling}. Recent advancements have introduced contact-implicit approaches, allowing for the dynamic determination of mode sequences \cite{fcimpc,posa2014direct}. While these methods improve constraint handling, they have notable drawbacks: dynamic feasibility is not guaranteed until the optimization is completed, and they often suffer from high computational complexity, which limits their real-time applicability.

Alternatively, recent studies have adapted indirect methods for hybrid systems \cite{Li2020,kong2021ilqr,Kong2023base}, which optimize state trajectories by indirectly optimizing the control sequence. These methods are computationally efficient and ensure dynamic feasibility throughout the optimization. However, existing approaches primarily focus on unconstrained scenarios, often overlooking additional constraints inherent in real-world applications, such as collision avoidance.

\begin{figure}[t!]
    \centerline{\includegraphics[width=\columnwidth]{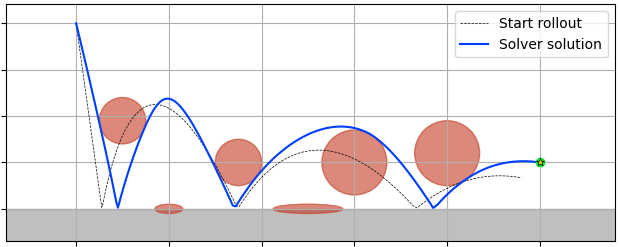}}
    \caption{Constrained hybrid iLQR computes optimal trajectories for hybrid dynamical systems, avoiding obstacles while respecting dynamic constraints.\label{fig1}}
    
\end{figure}

This work investigates the handling of input and external state constraints for hybrid dynamical systems within an indirect method. We aim to advance the capabilities of hybrid systems in diverse constrained scenarios while preserving computational efficiency and ensuring dynamic feasibility (Fig.~\ref{fig1}). The key contributions of our study include:
\begin{enumerate}
    \item Extending the recently proposed Hybrid iLQR method \cite{kong2021ilqr} with constraint handling capabilities \cite{almubarak2022safety, howell2019altro},
    \item Formulating two constraint handling mechanisms for hybrid setting: Discrete Barrier State and Augmented Lagrangian methods,
    \item A comparative simulation study of both methods that analyzes convergence performances and evaluates their efficiencies for infeasible starting trajectories.
    \item Open-sourcing the implementation of the proposed methods at \href{https://sites.google.com/view/c-hilqr}{https://sites.google.com/view/c-hilqr}.
\end{enumerate}

\section{Related Work}

Recent literature on trajectory optimization for hybrid dynamical systems can be mainly investigated through two primary approaches: direct and indirect methods.

\textbf{Direct methods} transcribe the Optimal Control Problem (OCP) into a Nonlinear Programming (NLP) problem, and have been widely employed in robotic trajectory planning and control. However, hybrid dynamics involving continuous dynamics and discrete events often necessitate Linear Complementarity Programs (LCP) or Mixed Integer Programs (MIP) \cite{posa2014direct, patel2019contact}. Pardo \textit{et al.} \cite{pardo2017hybrid} presented an algorithm for legged robots' locomotion, employing direct collocation within the constraint-consistent subspace defined by contact configurations. This method, incorporating projected impact dynamics constraints, was validated on a hydraulically-actuated quadruped robot. Posa \textit{et al.} \cite{posa2016optimization} introduced DIRCON, extending direct collocation with manifold constraints and third-order integration accuracy, demonstrating its efficacy in complex locomotion tasks. While these methods predefine a mode or gait sequence, limiting scalability, contact-implicit frameworks have improved optimization capabilities. Patel \textit{et al.} \cite{patel2019contact} improved trajectory accuracy by combining direct collocation using higher-order orthogonal polynomials with contact-implicit optimization, eliminating the need for mode scheduling and a priori contact knowledge. Posa \textit{et al.} \cite{posa2014direct} addressed trajectory planning with inelastic contact models formulated as LCP, solved via Sequential Quadratic Programming, achieving accurate contact-implicit trajectory optimization. Earlier, Yunt \cite{yunt2011augmented} proposed an Augmented Lagrangian-based direct shooting method for switching Lagrangian systems, eliminating the need for explicit mode scheduling by formulating the NLP in the mathematical programming with equilibrium constraints framework. Additionally, Manchester \textit{et al.} \cite{manchester2019contact} developed algorithms based on discrete variational mechanics for higher-order time-stepping methods, enhancing the accuracy of contact-implicit trajectory optimization.

\textbf{Indirect methods}, such as Differential Dynamic Programming (DDP), initially relied on predefined mode sequences, simplifying the optimization but limiting adaptability and optimality. Morimoto \textit{et al.} \cite{morimoto2002minimax} introduced a Differential Dynamic Programming (DDP) algorithm based on the minimax criterion, enhancing robustness against disturbances in high-dimensional state space robots. Budhiraja \textit{et al.} \cite{budhiraja2018differential} advanced this by developing a trajectory optimization algorithm for whole-body motion planning using Karush-Kuhn-Tucker (KKT) constraints, tested effectively on the HRP-2 robot. Li \textit{et al.} \cite{Li2020} attempted to map state perturbations through hybrid transitions using the Jacobian of the reset map, revealing limitations when the initial rollout did not include the optimal mode sequence. To overcome such limitations, Carius \textit{et al.} \cite{carius2018trajectory} developed a contact-invariant trajectory optimization algorithm, integrating bilevel optimization concepts within the unconstrained iLQR algorithm. Kong \textit{et al.} \cite{kong2021ilqr, Kong2023base} introduced the Hybrid iterative Linear Quadratic Regulator (HiLQR), using the saltation matrix \cite{kong2023saltation} to handle state deviations during hybrid transitions, achieving mode sequence optimization within the DDP framework. Their subsequent work expanded DDP into a Model Predictive Control (MPC) approach for real-time trajectory tracking.

In direct methods, handling external constraints is seamlessly integrated within the optimization framework, as demonstrated by tools like IPOPT \cite{wachter2006implementation} and SNOPT \cite{gill2005snopt}. However, indirect methods present a challenge in naturally accommodating these constraints. Various approaches have been explored to address this issue, particularly for Differential Dynamic Programming (DDP) methods. These include handling input constraints through techniques such as squashing functions and hybrid strategies \cite{tassa2014control,marti2020squash}, addressing state constraints with active set methods using Karush-Kuhn-Tucker (KKT) conditions \cite{xie2017differential}, and employing penalty methods like the Augmented Lagrangian TRajectory Optimizer (ALTRO) \cite{howell2019altro}. Additionally, primal-dual interior point methods \cite{pavlov2021interior} and Discrete Barrier State methods \cite{almubarak2022safety} have been developed to ensure trajectory safety. In the context of stochastic DDP, Linear Quadratic Gaussian (LQG) control has been used to manage input constraints \cite{todorov2005generalized}, while active set methods have been applied to enforce nonlinear safety constraints under system uncertainties \cite{Alcan2022}.

This work aims to advance trajectory optimization algorithms for hybrid systems using indirect methods by incorporating constraint-handling mechanisms inspired by those developed for smooth systems.

\section{Problem Definition}

We consider a hybrid dynamical system \cite{Kong2023base}, represented as a tuple
\begin{equation}
    \label{eq:hds}
    H := (\mathcal{J}, \Gamma, \mathcal{D}, \mathcal{F}, \mathcal{G}, \mathcal{R})
\end{equation}
where the parts are defined as:
\begin{itemize}
    \item $\mathcal{J} := \{I, J, \ldots, K\} \subset \mathbb{N}$ is the finite set of discrete \textit{modes}.
    \item $\Gamma \subset \mathcal{J} \times \mathcal{J}$ is the set of discrete \textit{transitions} that form a directed graph structure on $\mathcal{J}$.
    \item $\mathcal{D}$ is the collection of \textit{domains} $D_I$.
    \item $\mathcal{F}$ is a collection of time-varying vector fields $F_I$.
    \item $\mathcal{G}$ is the collection of \textit{guards} where $G(I,J)(t) = \{(x, t) \in D_I \mid g(I,J)(t, x) \leq 0\}$.
    \item $\mathcal{R}$ is called the \textit{reset} that maps the state from $D_I$ to $D_J$ when the guard $G(I,J)$ is met.
\end{itemize} 


Given the transition dynamics $x_{k+1} = f_{\Delta}(x_k,u_k,\Delta t)$, we account for hybrid transitions and changes in the dynamic component $f_{\Delta}$ by introducing the saltation matrix. The \textit{saltation matrix}, which characterizes the effect of a hybrid transition from domain $D_I$ to $D_J$, is defined as:
\begin{equation}
    \label{eq:salt-matr}
    \Xi :=  D_x R + \frac{(F_J - D_x R F_I - D_t R) D_x g}{D_t g + D_x g F_I}
\end{equation}
where 
\begin{equation*}
    \begin{split}
        & R := R_{(I,J)}(t^-, x(t^-), u(t^-)), g := g_{(I,J)}(t^-, x(t^-), u(t^-)),\\
        & F_I := F_I(t^-, x(t^-), u(t^-)), F_J := F_J(t^+, x(t^+), u(t^+))
    \end{split}        
\end{equation*}

The saltation matrix represents the first-order approximation of variations at hybrid transitions and maps perturbations from pre-transition $\delta x(t^-)$ to post-transition $\delta x(t^+)$ as follows: 
\begin{equation}
    \delta x(t^+) = \Xi_{I,J} \delta x(t^-) + \text{h.o.t.}
\end{equation}
The use of this formulation will allow us to linearize the system around hybrid transitions.



Given the definition of a hybrid dynamical system (\ref{eq:hds}), we state the problem as follows: For a given initial state $x_0$, a goal state $x_{goal}$, and a time horizon $N$, the aim is to find an input trajectory $(u_0,..., u_{N-1})$ that minimizes the objective function:
\begin{equation} \label{eq:trajopt}
\begin{aligned}
\min_{u_{0:N-1}} & \quad J = \ell_N(x_N) + \sum_{k=0}^{N-1}\ell_k(x_k,u_k,\Delta t)\\
\textrm{s.t.} & \quad x_{k+1} = f_{\Delta}(x_k,u_k,\Delta t), \quad k = 0, \ldots, N-1, \\
& \quad g_j(x_k,u_k) < 0, \quad \forall j,\\
& \quad u_{min} \leq u_k \leq u_{max},
\end{aligned}
\end{equation}
where $g$ represents the general nonlinear state constraints such as obstacle avoidance and state boundaries. $\ell$ and $\ell_N$ are running and final cost functions, respectively. In the context of our work, $f$ is not represented by a single function but rather by a collection of dynamic equations each one representing a different mode of the system.





\section{Hybrid iLQR}
\label{sec:hilqr}
The unconstrained version of the problem formulated in (\ref{eq:trajopt}) was the focus of the works by Kong et al. \cite{kong2021ilqr,Kong2023base}, where they were the first to adapt the smooth iLQR algorithm to hybrid systems with hard contact models using the saltation matrix. In this work, we extend their approach by formulating a unified constraint-handling mechanism for hybrid iLQR. We implement two different approaches: the safety-embedded Discrete Barrier State HiLQR, an interior point method, and the Augmented Lagrangian HiLQR, a penalty-based method.

Here we describe the fundamental steps of HiLQR for completeness. Similar to classical iterative LQR, the first rollout step of the algorithm involves running a forward simulation of the system dynamics using a control input sequence $\{u_0, ..., u_{N-1}\}$, which can either be randomly generated or predefined. Two main hybrid simulation techniques can be employed to forward simulate the system: event-driven hybrid simulators and time-stepping algorithms \cite{Kong2023base}. In this work, a time-stepping algorithm is used to integrate over small time steps, to avoid the Zeno problem \cite{lygeros2003dynamical} and to capture intermediate impacts that trigger \textit{mode changes} in the system dynamics. Each intermediate eventi is stored in a vector to be subsequently used in the backward pass.


Given the cost functions in (\ref{eq:trajopt}), the cost-to-go and action-value functions $V$ and $Q$ are defined as
\begin{equation}
\label{eq:Q_and_V}
\begin{aligned}
    V_k(x_k) & =\underset{u_k}{\min}\{\ell_k(x_k,u_k)+V_{k+1}(f(x_k,u_k))\} \\
    &=\underset{u_k}{\min}\;Q(x_k,u_k).
\end{aligned}
\end{equation}
where the action-value function $Q$ is approximated up to second order as:
\begin{equation}
\label{eq:Q-func}
\delta Q_k = \frac{1}{2} \begin{bmatrix}
    \delta x_k \\ \delta u_k
\end{bmatrix}^T \begin{bmatrix}
    Q_{xx} & Q_{xu} \\ Q_{ux} & Q_{uu}
\end{bmatrix} \begin{bmatrix}
    \delta x_k \\ \delta u_k
\end{bmatrix} + \begin{bmatrix}
    Q_{x} \\ Q_{u}
\end{bmatrix} \begin{bmatrix}
    \delta x_k \\ \delta u_k
\end{bmatrix}
\end{equation}
The block matrices follow the structure in \cite{Kong2023base}, including the \textit{saltation matrix} to map the state evolution through hybrid transitions as follows:
\begin{equation}\label{eq:HiLQR}
    \begin{split}
         Q_{x} &= \ell_{x}+ f_x\Tra\Xi\Tra V_x'\\
        Q_{u} &= \ell_{u}+ f_u\Tra\Xi\Tra V_x'\\
        Q_{xx} &= \ell_{xx} + f_x\Tra\Xi\Tra V_{xx}'\Xi f_x\\
        Q_{uu} &= \ell_{uu} + f_u\Tra\Xi\Tra V_{xx}'\Xi f_u\\
        Q_{xu} &= \ell_{xu} + f_u\Tra\Xi\Tra V_{xx}'\Xi f_x\\
    \end{split}
\end{equation}
where $V_{xx}'$ and $V_{x}'$ are the Hessian and gradient of the cost-to-go at next time step, respectively. Minimizing (\ref{eq:Q-func}) with respect to $\delta u_k$ results in the following affine controller:
\begin{equation}
    \delta u^* = -Q_{uu}^{-1}(Q_{ux}\delta x+Q_u) = K \delta x + d
\end{equation}

Subsequently, the optimal controller gains are used to calculate the derivatives of $V$ at current time-step:
\begin{equation}
\label{eq:V-derivatives}
    \begin{aligned}
        V_{x} & = Q_{x}+K Q_{u} + K\Tra Q_{uu}d+Q_{ux}\Tra d,\\
        V_{xx} & = Q_{xx}+K\Tra Q_{uu}K+K\Tra Q_{ux}+Q_{ux}\Tra K.
    \end{aligned}
\end{equation}

In the backward pass, all local optimal controller gains ($K,d$) are sequentially calculated in reverse, starting from the final state, where $V_N(x_N)=\ell_N(x_N)$ and progressing back to the initial state. In the forward pass, those controller gains are employed to update the control sequence starting from the initial state as follows:
\begin{equation}
        \begin{split}
        & \delta x_k = \bar{x}_k - x_k\\
        & \delta u_k = K_k \delta x_k + \alpha d_k\\
        & \bar{u}_k = u_k + \delta u_k
    \end{split}
\end{equation}
where $\alpha$ represents a line-search parameter. The backward and forward pass steps follow each other until the convergence is achieved. The convergence of iLQR methods is highly dependent on the search direction and step size. For detailed practices regarding regularization and line search, please refer to \cite{howell2019altro}.

In the following subsections, we will adapt two constraint-handling techniques to appropriately transform the optimization problem presented in (\ref{eq:trajopt}) into an unconstrained form, allowing it to be solved using the hybrid iLQR method.

\subsection{DBaS-HiLQR}

Discrete Barrier States (DBaS) \cite{almubarak2022safety} is a recently developed interior point algorithm for handling constraints in indirect optimization techniques for smooth dynamical systems. This approach integrates safety constraints into the system's dynamics and performance objectives. In this section, we outline the necessary steps of DBaS for clarity and present how to apply them to the constrained formulation of hybrid iLQR, which we refer to as DBaS-HiLQR.

To embed the constraints into system dynamics, the state vector is augmented as follows:
\begin{equation}
    \tilde{x}_k = \begin{bmatrix}
        x_k \\ w_k
    \end{bmatrix}
\end{equation}
where $w_k$ represents the barrier state, capturing state constraints:
\begin{equation}
    \label{eq:bar-state}
    w_k = B(g(x_k)) - B(g(x_{goal}))
\end{equation}
Here, $g(\cdot)$ denotes the inequality constraint function as defined in (\ref{eq:trajopt}), and $B(g(x_k))=-1/g(x_k)$ is chosen as the barrier function, known as the Carroll barrier, i.e., inverse barrier function, which has been proven to be sufficient for guaranteeing boundedness and safety \cite{agrawal2017discrete, almubarak2022safety}.

For smooth dynamical systems, the next barrier state $w_{k+1}$ is typically calculated by evaluating (\ref{eq:bar-state}) at $x_{k+1}$$=$$f_{\Delta}(x_k,u_k,\Delta t)$ \cite{almubarak2022safety}. For hybrid dynamical systems, due to the necessity of assessing transitions via the saltation matrix, we modify the barrier state dynamics by using its continuous-time derivative $\dot{w}(t)$ as follows:
\begin{equation}
    \dot{w}(x(t)) = \frac{1}{g(x(t))^2} g_x(x(t)) \, \dot{x}(t).
\end{equation}

The dynamics of the augmented states become the 
\begin{equation}
    \dot{\tilde{x}}(t) =   \tilde{f}(\tilde{x},t) = \begin{bmatrix}
        f(x(t),u(t))\\
        \dot{w}(x(t))
    \end{bmatrix}.
\end{equation}

To account for the extended state vector, we augment the objective function weight matrices $Q_k,Q_N$ respectively with $q_w,q_{w_N}$, representing respectively the terminal and running cost of the Discrete Barrier State. The cost function thus includes a quadratic term of the barrier state value.

The update formulas for backward pass follow the standard formulation in (\ref{eq:HiLQR}), including the saltation matrix of the augmented system $\tilde{\Xi}$.

In the forward pass, the algorithm must be able to discard trajectories that violate constraints and so in this step, we define the Discrete Barrier State as 
\begin{equation}
    w_k = \begin{cases}
            B(h(x_k)) & g(x_k) < 0 \\
            \infty & g(x_k) \geq 0.
            \end{cases}
\end{equation}
With this formulation, the step-size line search will automatically discard trajectories violating constraints due to their infinite cost. This interior point approach strictly enforces constraint equation satisfaction.



Embedding constraints into the state vector allows constraint handling directly in the saltation matrix, mapping their evolution through hybrid transitions. This approach converts safety constraints into hard constraints, ensuring better constraint handling properties during intermediate iterations. In contrast, the augmented Lagrangian method, discussed in the next section, incorporates constraints only into the objective function, excluding them from state evolution analysis in hybrid transitions.

\subsection{AL-HiLQR}

The Augmented Lagrangian (AL) method \cite{howell2019altro} is a classical penalty approach that incorporates constraints into the objective function and iteratively increases the penalty for violations or near-violations of these constraints. Additionally, it maintains estimates of the Lagrange multipliers associated with the constraints, enabling convergence to the optimal solution without the need for penalty terms to grow indefinitely. In this section, we introduce the necessary steps of AL and apply it to the constrained formulation of hybrid iLQR, which we refer to as AL-HiLQR.

Given the problem in (\ref{eq:trajopt}), the Lagrangian formulation consists of augmenting the objective function as follows:
\begin{equation}
\label{eq:lagrangian-fnc}
\begin{split}
    & \mathcal{L}_A = \mathcal{L}_N(x_N,\lambda_N,\mu_N) + \sum_{k=0}^{N-1}\mathcal{L}_k(x_k,u_k,\lambda_k,\mu_k)\\
    & = \ell_N(x_N) + (\lambda_N + \frac{1}{2}g_N(x_N)I_{\mu,N})\Tra g_N(x_N)\\
    & + \sum_{k=0}^{N-1}[\ell_k(x_k,u_k) + (\lambda_k + \frac{1}{2}g_k(x_k)I_{\mu,k})\Tra g_k(x_k)]
\end{split}
\end{equation}
where $\lambda$ are the Lagrangian multipliers, $I_{\mu}$ is a diagonal matrix defined as
\begin{equation}
    I_{\mu} = \begin{cases}
            0 & g_i(x) < 0 \land \lambda_i = 0, i\in \mathcal{I}\\
            \mu_i & otherwise
            \end{cases}
\end{equation}
and $\mu$ are the penalty terms. AL-HiLQR algorithm includes the following steps:

\begin{enumerate}
    \item Fix $\lambda$ and $\mu$ parameters constant and solve the unconstrained problem for objective function (\ref{eq:lagrangian-fnc}) using HiLQR algorithm described in the beginning of Section \ref{sec:hilqr} considering the modifications in the value function at the terminal state:
    \begin{equation}
    \begin{split}
        & V_x = (\ell_N)_x + ((g_N)_x)^T(\lambda+I_{\mu N}g_N)\\
        & V_{xx} = (\ell_N)_{xx} + ((g_N)_x)^TI_{\mu N}(g_N)_x\\
    \end{split}
    \end{equation}
    and in the derivatives of action-value function:
    \begin{equation}
    \begin{split}
        & Q_{x} = \ell_{x}+ f_x^T\Xi\Tra V_{x}' + g_x^T(\lambda + I_{\mu}g)\\
        & Q_{u} = \ell_{u}+ f_u^T\Xi\Tra V_{x}' + g_u^T(\lambda + I_{\mu}g)\\
        & Q_{xx} = \ell_{xx} + f_x^T\Xi\Tra V_{xx}'\Xi f_x + g_x^TI_{\mu}g_x\\
        & Q_{uu} = \ell_{uu} + f_u^T\Xi\Tra V_{xx}'\Xi f_u + g_u^TI_{\mu}g_u\\
        & Q_{xu} = \ell_{xu} + f_u^T\Xi\Tra V_{xx}'\Xi f_x + g_u^TI_{\mu}g_x\\
    \end{split}
    \end{equation}
    \item If the resultant trajectory is not met the requirements for constraint violations, update the Lagrangian multipliers and the penalty terms:
    \begin{equation}
        \begin{aligned}
            \lambda_i &= \max(0,\lambda_i + \mu_ig_i(x^*)) \\
            \mu &= \phi \mu
        \end{aligned}
    \end{equation}
    \item Repeat until the convergence is satisfied.
\end{enumerate}

Compared to the DBaS-HiLQR algorithm, this method allows constraint violation during intermediate iterations. The initial values for penalty terms $\mu$ allow tuning the solver response to initial infeasibility, while the parameter $\phi$ tunes the constraint convergence speed. Decreasing $\mu_0$ and $\phi$ gives the solver higher chances to reach constraint convergence but at a higher computational cost, while the opposite way, guarantees to speed up the process, potentially reducing the success rate of the solver \cite{howell2019altro,alcan2025constrained}.

\section{Experimental Results}
We define a hybrid dynamical system consisting of a 2D bouncing ball that will serve as a test bench for our methods in various challenging scenarios. The state vector $x \in \mathbb{R}^4$ consists of the two positions and velocities along the two main axis, while the control input vector $u \in \mathbb{R}^2$ consists of the two forces along those main axis. The system is modeled as a hybrid system with two modes: mode 1 corresponds to $\dot{z} < 0$ and mode 2 to $\dot{z} \geq 0$. During both modes, the ball follows ballistic dynamics defined by the following relation:
\begin{equation}
    f_{1,2}(x,u) = \begin{bmatrix}
        v_y,
        v_z,
        \frac{F_y}{m},
        \frac{F_z-mg}{m}
    \end{bmatrix}\Tra
\end{equation}
The guard functions are defined as $g_{1,2}(x,u) = -z$ and $g_{2,1}(x,u) = -\dot{z}$. The reset map for the transition from mode 2 to mode 1 is given by $R_{2,1} = [y,z,\dot{y},\dot{z}]$, and the reset map from mode 1 to mode 2 is $R_{1,2} = [y,z,\dot{y},-e\dot{z}]\Tra$, where $e$ is the coefficient of restitution during the impact, set to $e = 0.75$. 

To evaluate the performance of both the DBaS-HiLQR and AL-HiLQR methods, tests were conducted under various starting conditions and obstacle placements\footnote{The videos of the experiments can be found on the project website:\\ \href{https://sites.google.com/view/c-hilqr}{\texttt{https://sites.google.com/view/c-hilqr}}}. The solvers were assessed in scenarios starting with either feasible or infeasible initial rollout trajectories.
For each number of obstacles ranging from 1 to 10, 100 scenarios were generated, providing a solid basis for statistical analysis of robustness, success rate (the percentage of scenarios where convergence criteria were met), iteration count, and final trajectory cost.

\begin{figure}[t!]
    \centerline{\includegraphics[width=\columnwidth]{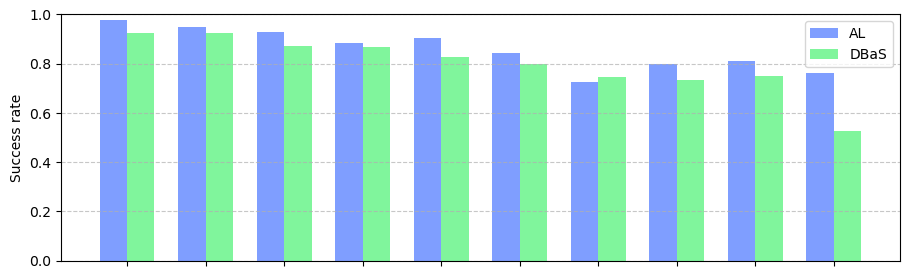}}
    \centerline{\includegraphics[width=\columnwidth]{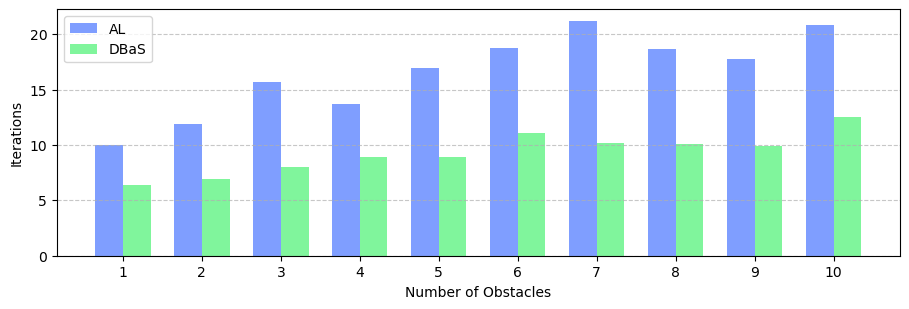}}
    \caption{Statistical comparison of DBaS-HiLQR and AL-HiLQR performance in randomly generated scenarios with feasible starting trajectories, in terms of success rates and convergence iterations.\label{fig:statistical-eval}}
\end{figure}

In all test cases, no reference trajectory was provided. The initial and goal positions were $x_0$$=$$ [0,4]\Tra$ and $x_{goal}$$=$$[10,1]\Tra$, respectively. The ball's mass was $m$$=$$1$ $kg$, gravity was $g$$=$$9.81$ $m/s^2$, simulated with $N$$=$$200$ and $\Delta t$$=$$0.02$ $sec$.

For both methods the weight matrices have been taken equal to $Q_T = 4\times10^{3}I_{2\times2}$ and $R = 5\times10^{-3}I_{2\times2}$. For the velocity component, the weights have been set to $0$. For the DBaS-HiLQR the barrier state weight has been chosen as $q_w = 1\times10^{-5}$. The solver archives convergence if all constraint tolerances are met and the error norm on the final position $\epsilon$ is $||\epsilon|| \leq 5\times10^{-2}$.


\subsection{Test Case 1: Feasible starting rollout trajectory}
In this test case, we assessed the methods using randomly generated scenarios with obstacles of various shapes, sizes, and positions. The initial rollout trajectories were created from randomly generated control input sequences, but only those that resulted in initially feasible state trajectories were considered.

Figure \ref{fig:statistical-eval} presents a statistical evaluation of all tests conducted for different numbers of obstacles (100 random trajectories for each number) in terms of success rate and iteration count. With a low number of obstacles ($\leq$6), both methods performed similarly, achieving success rates better than 80$\%$. However, as the number of obstacles increased, the AL approach outperformed the DBaS method due to its adjustable penalty mechanism. This mechanism allows the AL method to handle more complex scenarios but requires additional iterations. Consequently, in cases where both methods converged, the DBaS method typically required fewer iterations.

\begin{figure}[t!]
    \centerline{\includegraphics[width=\columnwidth]{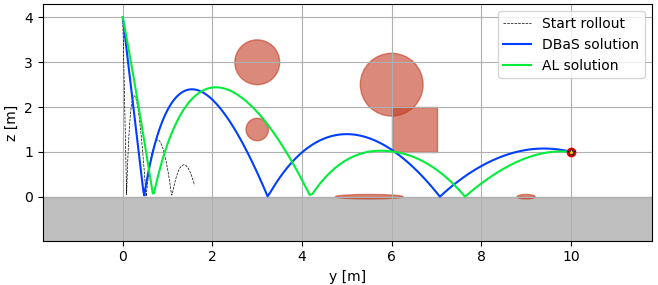}}
    \caption{Example trajectories generated by DBaS-HiLQR and AL-HiLQR, highlighting the conservative updates of DBaS-HiLQR versus the ability of AL-HiLQR to navigate tighter spaces. \label{fig:example-trajs}}
\end{figure}

A typical result of the trajectories generated by both methods is presented in Figure \ref{fig:example-trajs}. The DBaS method tends to make more conservative updates and maintains feasibility throughout the optimization process. In contrast, the AL method allows for temporary infeasibility, enabling the final trajectory to navigate through tighter spaces and meet more stringent constraints.

\begin{figure*}[t!]
    \centering
    \includegraphics[width=0.34\textwidth]{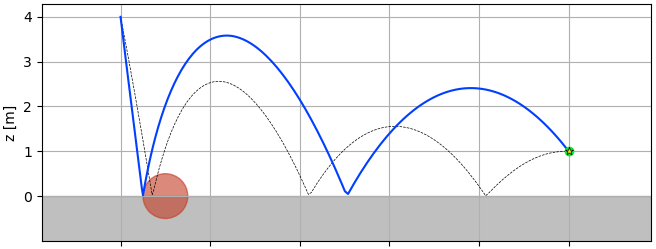} 
    \hfill
    \includegraphics[width=0.32\textwidth]{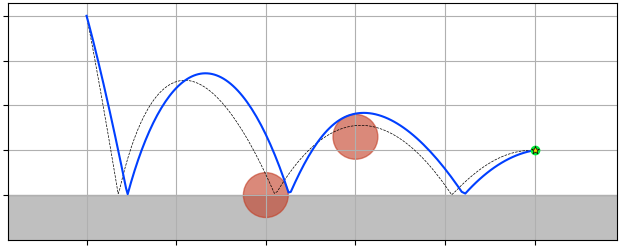} 
    \hfill
    \includegraphics[width=0.32\textwidth]{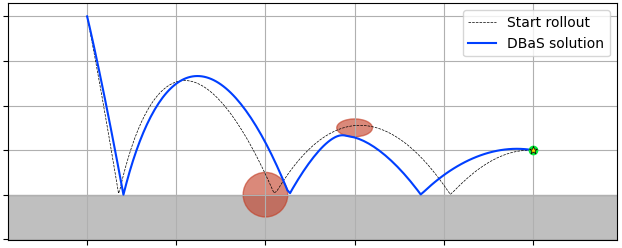} 
    
    \caption{Trajectories generated by the AL-HiLQR algorithm, demonstrating adaptation to obstacles introduced during hybrid transitions and showcasing the algorithm's ability to replan and adjust to new constraints.\label{fig:infeasible-tests}}
    
\end{figure*}

\subsection{Test Case 2: Infeasible Starting Rollout Trajectory}
In this test case, we evaluated how well the methods handle infeasible starting trajectories. The initial rollout trajectories were generated from randomly created control inputs, but only those resulting in infeasible state trajectories were considered.

\begin{table}[t!]
\centering
\caption{Average Success Rates Across All Experiments}
\begin{tabular}{c|c|c|c}
 & \textbf{Feasible Start} & \textbf{Infeasible Start} & \textbf{Difference ($\Delta$)} \\ \hline  
\textbf{DBaS-HiLQR} & 84.1\% & 61.9\% & -26.5\% \\ 
\textbf{AL-HiLQR} & 87.4\% & 69.4\% & -20.6\% \\ 
\end{tabular}
\label{tab:gray_table}
\end{table}

Table \ref{tab:gray_table} presents the average success rates across all experiments, considering all numbers of obstacles, and compares the performance of DBaS and AL in terms of feasible and infeasible starts. The results show that AL-HiLQR outperforms DBaS-HiLQR in handling infeasible starts, with only a $20.6\%$ drop in success rate, compared to $26.5\%$ for DBaS-HiLQR. AL-HiLQR performs better in handling infeasibilities due to its adaptability to additional constraints. In contrast, DBaS-HiLQR’s safety-embedded approach struggles when operating outside the safe set, leading to instability and a $23.4\%$ failure rate due to numerical issues.

We further tested AL-HiLQR by introducing constraints during mode transitions. Figure \ref{fig:infeasible-tests} shows that it successfully replans the contact sequence in these cases.

\section{Conclusion}
In this work, we developed a novel constrained trajectory optimization algorithm for hybrid dynamical systems by extending the Hybrid iLQR method to handle external state and input constraints within an indirect optimization framework. By integrating the Discrete Barrier State and Augmented Lagrangian methods into the HiLQR algorithm, we enhanced its capabilities to address complex constraints such as obstacle avoidance, state boundaries, and actuation limits.
Implementing and testing the proposed algorithms on real-world platforms to evaluate their real-time performance and scalability remains an open issue. Additionally, the integration of learning-based approaches to enhance the adaptability and efficiency of constraint handling in hybrid systems presents an intriguing opportunity for further studies.


\bibliography{mybib}{}
\bibliographystyle{ieeetr}

\end{document}